\documentstyle[aps,epsf,pre,twocolumn]{revtex}

\draft 

\begin{document}

\title{Modulation of Localized States in Electroconvection}

\author{Carina Kamaga}
\author{Michael Dennin}
\address{Department of Physics and Astronomy}
\address{University of California at Irvine}
\address{Irvine, CA 92697-4575.}

\date{\today}

\maketitle

\begin{abstract}

We report on the effects of temporal modulation of the
driving force on a particular class of localized states, known as worms,
that have been observed in electroconvection in nematic liquid crystals.
The worms consist of the superposition of traveling waves and have been
observed to have unique, small widths, but to vary in length. The transition
from the pure conduction state to worms occurs via a
backward bifurcation. A possible explanation of the formation of the worms
has been given in terms of coupled amplitude equations. Because the worms
consist of the superposition of traveling waves, temporal modulation of
the control parameter is a useful probe of the dynamics of the system.
We observe that temporal modulation
increases the average length of the worms and stabilizes worms below the
transition point in the absence of modulation.
\end{abstract}


Patterns in spatially extended dissipative systems have
been of interest in a wide range of disciplines. Their
formation is associated with nonlinear and nonequilibrium effects
that produce qualitatively new phenomena that are not observable in linear
systems \cite{REV}. One of the more interesting classes of patterns is localized states,
or pulses. These states correspond to the coexistence of small regions in
which a pattern exists with larger regions of the uniform state. Interest
in localized states has grown dramatically since the experimental discovery
of pulses in
Rayleigh-B\'{e}nard convection using binary fluids in narrow channels \cite{MFS87,HAC87}.
These quasi-one dimensional states have been described theoretically in a number
of ways \cite{TF88,R96,MN90,HP91,HR95}. Other quasi-one dimensional structures
have been observed in Taylor vortex flow \cite{WM92,GS97}, directional
solidification \cite{SBL88}, cellular flames \cite{BMR94}, and models of
parametrically driven waves \cite{GR96}. In two dimensions, localization
appears to be harder to achieve. In binary fluids, only long-lived
states have been observed \cite{LBCA93}. Recently,
truly stable states have been observed in
granular materials \cite{UMS96}, viscous fluids \cite{LAF96}, and
electroconvection \cite{JR88,DAC96,BFFPC97}.
In this paper, we will focus
specifically on the localized states, known as worms \cite{DAC96}, that have been
observed in electroconvection in nematic liquid crystals.

Worms are unique for two reasons. First, worms occur in a system that is
intrinsically anisotropic: electroconvection in
nematic liquid crystals \cite{SMREFS,ECART}.
A nematic liquid crystal \cite{LC} is composed of
long rod-like molecules and has an inherent orientational order. The
axis parallel to the average alignment of the molecule is called the
director. For electroconvection, a nematic liquid crystal is placed
between two glass plates that have been treated to produce uniform
alignment of the director. This selects an axis, which we will refer
to as the x-axis. An ac voltage of rms amplitude $V$ and frequency $f$ is
applied across the two glass plates using transparent conductors.  
Below a well-defined value of the applied voltage, $V_o$, the system
is uniform. Above $V_o$, a pattern develops. For a relatively wide
range of parameters, the initial pattern consists of a collection of
worms. Worms consist of a superposition of traveling roll states
localized within an envelope. The wavevector of the rolls has a nonzero
angle with respect to the x-axis. Such states are referred to as oblique
rolls. Because the director only defines an axis, rolls with angle
$\theta$ and $\pi - \theta$ are degenerate and referred to as zig and
zag rolls. Therefore, there are four possible states that combine to
form worms: right- and left-traveling zig and zag rolls.

Before discussing the second feature that makes worms unique, we will
briefly review their known properties \cite{DAC96,BA98}.
The envelope of the worms
has a well-defined width in a direction approximately
perpendicular to the undistorted director. However, its
length is found to be irregular for values of
$V$ close to the onset voltage $V_o$. As $V$ is increased, the average length of
the worms increases until the worms extend across the system. In addition,
the vertical spacing between worms decreases, until the system is filled
with convection. The worms travel in a direction opposite the direction
of travel of the rolls that comprise the worms.
Other localized states, described as ``bursting'', were also
observed in a different parameter range. These states did not have
a well-defined width. The onset of worms occurs well below the
supercritical transition to the extended state, i.e. it is a subcritical
transition \cite{BA98}. It is this fact that is the second unique feature
of worms.

Most localized states occur because
the system exhibits bistability between an extended wave state and
the uniform state. This fact is essential to most theoretical
models of localized structures (e.g. \cite{MN90,HP91,HR95,BSC88,SB96,T97,CR99}).
Even though the worms occur {\it subcritically},
because the extended state occurs via a {\it supercritical} bifurcation,
one can not describe
the system in terms of fronts connecting the basic and nonlinear states,
ruling out the above mentioned mechanisms. 
A set of coupled amplitude equations have been proposed to explain the
worms \cite{RG98}. These consist of amplitude equations for the zig and zag modes that
are coupled to an additional weakly damped scalar mode.
Simulations of the amplitude equations
exhibit solutions that have the same general features as the worm state
and the bursting states. However, the additional slow mode has not yet 
been identified for electroconvection. 

In this paper, we report on an additional experimental
test of the proposed amplitude equation 
model. Because the worms consist of a superposition of traveling waves,
a temporal modulation of the driving voltage can couple the different
right- and left-traveling modes and produce interesting effects,
such as standing waves (e.g. \cite{RCK88,W88,RRFJS88,JZR89,JR90}).
For extended states, the strongest coupling occurs for a modulation
frequency at twice the natural traveling frequency. We expect the same
to be true for the worm state, so in this
paper, we focus on the 2:1 resonance.
We observe both a stabilization of the worms below their onset in the absence
of modulation and an increase in their natural length.

We used the nematic liquid crystal
{\it 4-ethyl-2-fluoro-4$^{\prime}$-[2[(trans-4-pentylcyclohexyl)-ethyl] byphenyl}
(I52) \cite{FGWP89}
for the experiments reported here. The liquid crystal was doped with
8\% by weight molecular iodine. A commercial cell was used for the
experiments \cite{EHCO}.
It was made from two glass slides that were coated with a transparent conductor,
a layer of indium-tin oxide (ITO).  The conductive coating was etched to form a
0.5 cm x 0.5 cm square electrode in the center of a 2.5 cm x 2.5 cm cell.
The glass slides were coated with a rubbed polyimide to align the director.

The sample was held in an aluminum block that provided temperature control.
The temperature was kept constant to $\pm 0.005 ^{\circ}{\rm C}$. The block
had glass windows in the bottom and top. The patterns were imaged by shining
light from below the sample and observing it from above with the standard
shadowgraph technique \cite{RHWR89}.

The temperature was varied over the range
$40 ^{\circ}{\rm C}$ to $60 ^{\circ}{\rm C}$. This produced a shift in the
onset voltage $V_o$ of the worms and the traveling frequency of the underlying
rolls due to changes in the material parameters, such as the electrical
conductivity.  
The onset voltage was measured by stepping the voltage in steps of 0.1 V
at a fixed applied frequency and waiting for 5 minutes at each step. The
onset voltage was defined to be the voltage at which worms were first
observed to exist. Recall, this is different than the actual critical
voltage, as the transition is backward with a large hysteresis. At 
temperatures lower than $40 ^{\circ}{\rm C}$, no worms were observed.

For the modulation experiments, the applied voltage has the form
$V(t) = \sqrt{2}[V + V_m cos(\omega_m t)]cos(\omega t)$.  In this paper,
$V_o$ is the measured onset
voltage for the worms in the absence of modulation ($V_m = 0$).
In this system of electroconvection at larger values of the
conductivity, the initial transition is to an extended state \cite{DAC96b}.
For this case, a large enough value of $V_m$ produces standing
waves that consist of only zig or zag rolls \cite{D00}. Therefore, it was
expected that the modulation would stabilize ``standing worms'';
however, it was unknown whether or not such objects would be stable.

For our system, the worm envelope did not travel significantly, as
previously observed \cite{DAC96}. This was due to the fact that
our worms are a superposition of both left- and right-traveling
rolls. Therefore, they were ``standing worms''. The exact type
of worm, traveling or standing, depends on the various material
parameters in a currently unknown fashion, so it
is not surprising that standing worms are observed in our system.
Further work is needed on this issue, but standing worms have been
reported for other samples \cite{DENNINTHESIS}.
The worms blinked with a frequency of
0.5 Hz to 2.5 Hz, depending on the parameters. Figure 1 shows 
images of a standing worm at four different times.  All these images were taken
at the temperature $T = 55 \pm 0.02 ^{\circ}{\rm C}$ and at the same location in
the cell.  The images of Fig. 1(a) through 1(c) were taken 0.2 s apart and
demonstrate the standing nature of the worms. 
Figure 1(d) shows the worm 1 hour later and demonstrates both the stability of
the worm and the fact that it does not travel any significant distance.
Because of the narrow width of the worms and nonlinear effects in the shadowgraph,
it is difficult to determine if these standing worms are a superposition of all four
modes (right- and left-traveling zig and zag) or simply two modes (right- and
left-traveling either zig or zag). The rolls in the images
in Fig. 1 are at a slight angle, suggesting
the superposition of two modes. (If both zig and zag rolls are present, the nonlinear
shadowgraph effects give the appearance of normal rolls \cite{DAC96}.) In fact,
most of the images are consistent with the superposition of only two modes; however,
some of the worms appear to be comprised of all four (see for instance, Fig. 2b).
The exact nature of the superposition is important for
comparison with the amplitude equation
predictions. One must use parameters in the amplitude equations that
correspond to type of standing worms present in the experiment.
At this point, the various types of standing worms have not been studied
in any detail theoretically, so we are not able to make any quantitative
comparisons.

Figure 2 compares an example of a worm state with $V_m = 0$ and
$V_m \neq 0$. For both cases, the onset voltage was 22.5 V, and
the temperature was set to $56 \pm 0.02 ^{\circ}{\rm C}$.  The
image on the left (a) was taken without the modulation, and the one
of the right (b) was taken with modulation. Without the modulation,
the worms are relatively short, consisting of only 3-4 convection
rolls inside. With the modulation, the worms grow substantially in
length, but retain essentially the same width. 

Figure 3 and 4 provide a more detailed comparison between the modulated
and unmodulated states. The images were taken at
$T = 60 ^{\circ}{\rm C}$ with an applied frequency of 50 Hz.
Without modulation, the onset of worms was found to be at 21.59 V, and the
traveling frequency of rolls inside the worm envelope was found to be 
$f = 2.08\, {\rm Hz}$.
The images (a) - (d) in Fig. 3 show the transition from below onset to above
onset. In particular, the length of the worms in image (b) is still
relatively short. As the applied voltage is increased, the number of worms 
increases, and the intensity of worms becomes stronger. 
The images in Fig. 4 (a) - (d) show the similar range of applied
voltages as in Fig. 3 (a) - (d), but now the applied voltage has a 
modulation voltage of $2.0\, {\rm V}$ and a modulation frequency 
$f_m = 4.166\, {\rm Hz}$. Images (a), (b) and (c) show that worms can be 
stabilized below the onset voltage in
the absence of modulation. Also, the modulation clearly generates
extremely long straight worms (image c). These worms are standing
worms with a frequency of $2.08\, {\rm Hz}$. This confirms that the
system is responding at the 2:1 resonance, as discussed
in the introduction. Finally, at
high enough values of applied voltage, for a given modulation, the
worms fill the system and produce extended convection.

We have shown that temporal modulation increases the length of the
worms without affecting the width. Also, as expected from the
temporal modulation of extended states \cite{RCK88,W88,RRFJS88,JZR89,JR90},
worms can be resonantly
excited below onset.  The transition from the worms to the extended
convection appears to be qualitatively different in the modulated and
unmodulated case. In the unmodulated case, the extended state appears
to fill in the system from in between the worms. In the modulated case,
the worms appear to fill the system with an increasing density and
eventually lose their localization perpendicular to the director as they
merge into an extended state. However, more work needs to be done on this
transition to quantify it. Also, it is known from the theoretical work on traveling
worms, that is is possible for modulation to increase the length of the worms.
However, as mentioned, the standing worms have not been studied in any
detail theoretically, so quantitative comparisons with theory will be the
subject of future work.

\acknowledgments

We thank Hermann Riecke and Catherine Crawford for useful discussions. This work was
supported by NSF grant DMR-9975479. M. Dennin also thanks the
Research Corporation Cottrell Scholar and Sloan Fellowship for additional
funding for this work.

\begin{figure}
\caption{ A series of images taken at a temperature of $T = 55\,^{\circ}{\rm C}$,
a voltage of $V = 18.59\, {\rm V}$, and a frequency $f = 25\, {\rm Hz}$. Images (a)-(c) 
were taken 0.2 s apart, and the image (d) was taken 1 hr after at the same location 
of the cell. This shows that the worms do not travel much, 
and that they blink. The bar in (a) corresponds to 0.15 mm.}
\end{figure}

\begin{figure}
\caption{This set of two images shows a comparison of unmodulated  
and modulated worms.  Both images are taken at a temperature of 
$T = 56\,^{\circ}{\rm C}$.  The image on the left shows worms with the applied
voltage, $V = 22.5\, {\rm V}$ and the applied frequency, $f = 50\, {\rm Hz}$ in the
absence of modulation ($V_m = 0$).  The image on the right shows the modulated 
worms taken with the same
applied voltage and the applied frequency with the one on the left, and with
the modulation voltage, $V_m = 2.5\, {\rm V}$ and the modulation frequency of $f_m = 
3.846\,{\rm Hz}$. The bar corresponds to 0.15 mm.}
\end{figure}

\begin{figure}
\caption{These images are 
taken at a temperature of $T = 60\,^{\circ}{\rm C}$, with an applied frequency of
$f = 50\,{\rm Hz}$ in the absence of modulation.  For this experiment, the 
applied voltage was stepped up by 0.1 V.  These images were taken at an applied
voltage of (a)$V = 21.3\, {\rm V}$ (b)$V = 21.7 \, {\rm V}$ (c)
$V = 22.1 \, {\rm V}$ and (d)$V = 23.0\, {\rm V}$.
The bar in (a) corresponds to 0.15 mm.}
\end{figure}

\begin{figure}
\caption{These images illustrate the development of modulated worms by stepping 
down the applied voltage with a fixed modulated voltage of $V_m = 2.0\, {\rm V}$, 
modulated frequency, $f_m = 4.16\,{\rm Hz}$, and the applied frequency of 
$f_o = 50.0\, {\rm Hz}$.  The range 
of applied voltage is similar to that shown in Figure 3.
The images were taken at an applied voltage of (a) $V = 20.7\, {\rm V}$
(b)  $V = 20.9\,V$
(c)  $V = 21.1\, {\rm V}$ and (d) $V = 21.7\, {\rm V}$.
The bar in (a) corresponds to 0.15 mm.} 
\end{figure}


\begin{references}

\bibitem{REV} For reviews of pattern formation, see
M. C. Cross and P. C. Hohenberg, Rev. Mod. Phys. {\bf 65},
851 (1993), and J. P. Gollub and J. S. Langer, Rev. Mod. Phys.
{\bf 71}, s396 (1999).

\bibitem{MFS87} E. Moses, J. Fineberg, and V. Steinberg, Phys. Rev. A
{\bf 35}, 2757 (1987).

\bibitem{HAC87} R. Heinrichs, G. Ahlers, and D. S. Cannell, Phys. Rev. A
{\bf 35}, 2761 (1987).

\bibitem{TF88} O. Thual and S. Fauve, J. Phys. (Paris) {\bf 49}, 1829 (1988).

\bibitem{R96} H. Riecke, Phys. Rev. Lett. {\bf 68}, 301 (1992).

\bibitem{MN90} B. A. Malomed and A. A. Nepomnyashchy, Phys. Rev. A {\bf 42}, 6009 (1990).

\bibitem{HP91} V. Hakim and Y. Pomeau, Eur. J. Mech. B Suppl. {\bf 10}, 137 (1991).

\bibitem{HR95} H. Herrero and H. Riecke, Physica (Amsterdam) {\bf 85D}, 79 (1995).

\bibitem{WM92} R. J. Wiener and D. F. McAlister, Phys. Rev. Lett. {\bf 69}, 2915 (1992).

\bibitem{GS97} A. Groisman and V. Steinberg, Phys. Rev. Lett. {\bf 78}, 1460 (1997).

\bibitem{SBL88} A. J. Simon, J. Bechhoefer, and A. Libchaber, Phys. Rev. Lett. {\bf 61},
2574 (1988).

\bibitem{BMR94} A. Bayliss, B. Matkowsky, and H. Riecke, Physica (Amsterdam) {\bf 74D},
1 (1994).

\bibitem{GR96} G. D. Granzow and H. Riecke, Phys. Rev. Lett. {\bf 77}, 2451 (1996).

\bibitem{LBCA93} K. Lerman, E. Bodenschatz, D. S. Cannell, and G. Ahlers,
Phys. Rev. Lett. {\bf 70}, 3572 (1993).

\bibitem{UMS96} P. Umbanhowar, F. Melo, and H. Swinney, Nature (London) {\bf 382},
5461 (1998).

\bibitem{LAF96} O. Lioubashevski, H. Arbell, and J. Fineberg, Phys. Rev. Lett {\bf 76},
3959 (1996).

\bibitem{JR88} A. Joets and R. Ribotta, Phys. Rev. Lett. {\bf 60}, 2164 (1988).

\bibitem{DAC96} M. Dennin, G. Ahlers, and D. S. Cannell, Phys. Rev. Lett. {\bf 77},
2475 (1996).

\bibitem{BFFPC97} H. R. Brand, C. Fradin, P. L. Finn, W. Pesch, and P. E. Cladis,
Physics Letters A {\bf 235}, 508 (1999).

\bibitem{SMREFS} E. Bodenschatz, W. Zimmermann, and L.
Kramer, J. Phys. (France) {\bf 49}, 1875 (1988);
L. Kramer, E. Bodenschatz, W.  Pesch, W. Thom and W.
Zimmermann, Liquid Cryst. {\bf 5}, 699 (1989).

\bibitem{ECART} Review articles on electroconvection
can be found in I. Rehberg,
B. L. Winkler, M. de la Torre Ju\'{a}rez, S. Rasenat, and
W. Sch{\"{o}}pf, Festk{\"{o}}rperprobleme {\bf 29}, 35 (1989);
S. Kai and W. Zimmermann, Prog. Theor. Phys. Suppl. {\bf 99},
458 (1989); and L. Kramer and W. Pesch, Annu. Rev. Fluid
Mec. {\bf 27}, 515 (1995).

\bibitem{LC} P. G. de Gennes, {\it The Physics of Liquid Crystals}
(Claredon Press, Oxford, 1974); S. Chandrasekhar, {\it Liquid Crystals}
(Cambridge University Press, Cambridge, England, 1992).

\bibitem{BA98} U. Bisang and G. Ahlers, Phys. Rev. Lett. {\bf 80},
3061 (1998).

\bibitem{BSC88} D. Bensimon, B. I. Shraiman, and V. Croquette, Phys. Rev.
A {\bf 38}, 5461 (1988).

\bibitem{SB96} H. Sakaguchi and H. Brand, Physica (Amsterdam) {\bf 97D},
274 (1996).

\bibitem{T97} Y. Tu, Phys. Rev. E {\bf 56}, R3765 (1997).

\bibitem{CR99} C. Crawford and H. Riecke, Physica D {\bf 129}, 83 (1999).

\bibitem{RG98} H. Riecke and G. D. Granzow, Phys. Rev. Lett. {\bf 81}, 333 (1998).

\bibitem{RCK88} H. Riecke, J. D. Crawford, and E. Knobloch,
Phys. Rev. Lett. {\bf 61}, 1942 (1988).

\bibitem{W88} D. Walgraef, Europhys. Lett. {\bf 7}, 485
(1988).

\bibitem{RRFJS88} I. Rehberg, S. Rasenat, J. Fineberg,
M. de la Torre Ju\'{a}rez, and V. Steinberg, Phys. Rev.
Lett. {\bf 61}, 2449 (1988).

\bibitem{JZR89} M. de la Torre Ju\'{a}rez, W. Zimmermann, and
I. Rehberg, in {\it Nonlinear Evolution of Spatio-Temporal
Structures in Dissipative Continuous Systems}, F. H. Busse
and L. Kramer, eds., NATO ASI Series B: Physics Vol. 225
(Plenum Press, New York, 1990).

\bibitem{JR90} M. de la Torre Ju\'{a}rez and
I. Rehberg, Phys. Rev. A {\bf 42}, 2096 (1990).

\bibitem{RSK94} H. Riecke, M. Silber, and L. Kramer,
Phys. Rev. E {\bf 49}, 4100 (1994).

\bibitem{FGWP89} U. Finkenzeller, T. Geelhaar, G. Weber, and L. Pohl,
Liquid Crystals {\bf 5}, 313 (1989).

\bibitem{EHCO} E.H.C. CO., Ltd., 1164 Hino, Hino-shi, Tokyo, Japan.

\bibitem{RHWR89} S. Rasenat, G. Hartung, B. L. Winkler,
and I. Rehberg, Experiments in Fluids {\bf 7}, 412 (1989).

\bibitem{DAC96b} M. Dennin, G. Ahlers, and D. S. Cannell, Science
{\bf 272}, 388 (1996).

\bibitem{D00} M. Dennin, Phys. Rev. E {\bf 62}, 7842 (2000).

\bibitem{DENNINTHESIS} M. Dennin, Ph. D. Thesis (1995).

\end{references}
\end{document}